\documentclass[reprint,
superscriptaddress,showpacs,preprintnumbers,
nofootinbib,amsmath,amssymb, aps, prd,longbibliography]{revtex4-2}

\usepackage[utf8]{inputenc}
\usepackage{mathrsfs}
\usepackage{bm}
\usepackage{url}
\usepackage[normalem]{ulem}
\usepackage{mathtools}
\usepackage{array}
\newcolumntype{P}[1]{>{\centering\arraybackslash}p{#1}}
\newcolumntype{M}[1]{>{\centering\arraybackslash}m{#1}}
\usepackage[caption=false]{subfig}
\usepackage{dcolumn}
\usepackage{graphicx, epsfig}
\usepackage[dvipsnames]{xcolor}
\usepackage{mathrsfs}
\usepackage{bm}
\usepackage{yhmath}
\usepackage[caption=false]{subfig}
\usepackage[normalem]{ulem}
\usepackage{mathtools}
\usepackage{bigints}
\usepackage{float}
\usepackage[colorlinks = true, linkcolor = purple, urlcolor  = blue, citecolor = blue, anchorcolor = blue]{hyperref}
\usepackage{float}
\usepackage{multirow}
\usepackage{tikz,xcolor,hyperref}
\definecolor{darkgreen}{rgb}{0.0, 0.2, 0.13}
\definecolor{bostonuniversityred}{rgb}{0.8, 0.0, 0.0}
\definecolor{lime}{HTML}{A6CE39}
\DeclareRobustCommand{\orcidicon}{
	\begin{tikzpicture}
	\draw[lime, fill=lime] (0,0) 
	circle [radius=0.16] 
	node[white] {{\fontfamily{qag}\selectfont \tiny ID}};
	\draw[white, fill=white] (-0.0625,0.095) 
	circle [radius=0.007];
	\end{tikzpicture}
	\hspace{-2mm}
}
\foreach \x in {A, ..., Z}{\expandafter\xdef\csname orcid\x\endcsname{\noexpand\href{https://orcid.org/\csname orcidauthor\x\endcsname}
			{\noexpand\orcidicon}}
}
\def\t13{\mathrel{{\theta_{13}}}}
\def\y12{\mathrel{{\tan^2 \theta_{12}}}}
\def\c2{\mathrel{{\chi^2 }}}

\newcommand{\be}{\begin{equation}}
\newcommand{\ee}{\end{equation}}
\newcommand{\ba}{\begin{eqnarray}}
\newcommand{\ea}{\end{eqnarray}}

\begin{document}
\title{Ultrahigh-energy neutrino searches using next-generation gravitational wave detectors at radio neutrino detectors: GRAND, IceCube-Gen2 Radio, and RNO-G}
\author{Mainak Mukhopadhyay\hspace{-1mm}\orcidA{}}
\email{mkm7190@psu.edu}
\affiliation{Department of Physics; Department of Astronomy \& Astrophysics; Center for Multimessenger Astrophysics, Institute for Gravitation and the Cosmos, The Pennsylvania State University, University Park, PA 16802, USA
}
\author{Kumiko Kotera\hspace{-1mm}\orcidB{}}
\affiliation{Sorbonne Universit\'e, CNRS, UMR 7095, Institut d’Astrophysique de Paris, 98 bis bd Arago, 75014 Paris, France
}
\affiliation{Astrophysical Institute, Vrije Universiteit Brussels, Pleinlaan 2, 1050 Brussels, Belgium
}
\affiliation{Department of Physics; Department of Astronomy \& Astrophysics; Center for Multimessenger Astrophysics, Institute for Gravitation and the Cosmos, The Pennsylvania State University, University Park, PA 16802, USA
}
\author{Stephanie Wissel\hspace{-1mm}\orcidC{}}
\affiliation{Department of Physics; Department of Astronomy \& Astrophysics; Center for Multimessenger Astrophysics, Institute for Gravitation and the Cosmos, The Pennsylvania State University, University Park, PA 16802, USA
}
\author{Kohta Murase\hspace{-1mm}\orcidE{}}
\affiliation{Department of Physics; Department of Astronomy \& Astrophysics; Center for Multimessenger Astrophysics, Institute for Gravitation and the Cosmos, The Pennsylvania State University, University Park, PA 16802, USA
}
\affiliation{Center for Gravitational Physics and Quantum Information, Yukawa Institute for Theoretical Physics, Kyoto, Kyoto 606-8502 Japan}
\author{Shigeo S. Kimura\hspace{-1mm}\orcidD{}}%
\affiliation{Frontier Research Institute for Interdisciplinary Sciences; Astronomical Institute, Graduate School of Science, Tohoku University, Sendai 980-8578, Japan
}
\date{\today}
\begin{abstract}
Binary neutron star (BNS) mergers can be sources of ultrahigh-energy (UHE) cosmic rays and potential emitters of UHE neutrinos. The upcoming and current radio neutrino detectors like the Giant Radio Array for Neutrino Detection (GRAND), IceCube-Gen2 Radio, and the Radio Neutrino Observatory in Greenland (RNO-G) are projected to reach the required sensitivities to search for these neutrinos. In particular, in conjunction with the next-generation of gravitational wave (GW) detectors like Cosmic Explorer (CE) and Einstein Telescope (ET), GW-triggered stacking searches can be performed with the UHE neutrino detectors. In this work, we explore the prospects of such searches by implementing in our analysis an upper distance limit based on the sky-localization capabilities of the GW detectors from which meaningful triggers can be collected. We find that if each GW burst is associated with a total isotropic-equivalent energy of $\sim 10^{50} - 10^{51}$ erg emitted in UHE neutrinos, along with a corresponding beaming fraction of $1$\%, GRAND and IceCube-Gen2 Radio have a large probability ($\sim 99$\%) to detect a coincident neutrino event using the joint combination of CE+ET in a timescale of less than 15 years of operation for our fiducial choice of parameters. In case of nondetections, the parameter spaces can be constrained at $3\sigma$ level in similar timescales of operation. We also highlight and discuss the prospects of such joint radio neutrino detector network, their importance, and their role in facilitating synergic GW and neutrino observations in the next era of multimessenger astrophysics.
\end{abstract}
\maketitle 
\section{Introduction}
\label{sec:intro}
Cosmic accelerators provide us with a window to high-energy processes in the Universe. Our understanding of these violent phenomena is based on observations and evidences presented by the various messengers from these sources, namely, neutrinos, gravitational waves (GWs), photons, and cosmic rays (CRs). GW170817 detected in gravitational wave~\cite{LIGOScientific:2017vwq} and electromagnetic channels~\cite{Goldstein:2017mmi,LIGOScientific:2017zic,LIGOScientific:2017ync,DES:2017kbs,Coulter:2017wya,J-GEM:2017tyx,Valenti:2017ngx,Lipunov:2017dwd} serves as a prime example of the power of multimessenger astronomy. With the advent of improved detectors across all the messengers, the next era of multimessenger astronomy promises to be revolutionary in our understanding of the Universe at the highest energy scales.

In particular, neutrinos serve as one of the most prominent messengers in the context of delivering intrinsic information from their sources. The undeflected signatures provided by neutrinos play a key role in performing precision time-domain astronomy. The first evidence of high-energy astrophysical neutrinos from cosmic accelerators were presented by the the IceCube Collaboration in 2013~\cite{IceCube:2013cdw,IceCube:2013low}. Since then, evidences for high-energy neutrino emission have been found from multiple sources like NGC 1068~\cite{IceCube:2022der} and the Galactic plane~\cite{IceCube:2023ame}, which have also been observed in the electromagnetic channel.

Binary neutron star (BNS) mergers are known to be rich GW sources (e.g., Refs.~\cite{Shibata:2002jb,LIGOScientific:2017vwq,Radice:2016rys,Shibata:2017xht,PhysRevLett.120.111101,Shibata:2019wef,Radice:2020ddv}).
BNS mergers may have relativistic jets leading to short gamma-ray bursts (GRBs) and merger ejecta accompanied by shocks, which can be sources of very and ultrahigh-energy cosmic rays (UHECRs)~\cite{Takami:2013rza,Kimura:2018ggg,Rodrigues:2018bjg,Murase:2018utn,Farrar:2024zsm,Zhang:2024sjp}. 
High-energy neutrinos can be produced by jets and winds~\cite{Kimura:2017kan,Biehl:2017qen,Ahlers:2019fwz,Kimura:2018vvz,Decoene:2019eux}.
Alternatively, analogous to pulsars from collapsars, magnetars from low-mass BNS mergers could be UHECR and neutrino emitters~\cite{Murase:2009pg,Kotera:2011vs,Fang:2013vla,Lemoine:2014ala,Kotera:2015pya,Fang:2017tla,Carpio:2020wzg,Farrar:2024zsm,Mukho_nu}. 
A neutrino and GW channels have jointly not yet yielded any significant results. However, even though no neutrinos were observed from GW170817~\cite{ANTARES:2017bia,Super-Kamiokande:2018dbf,IceCube:2020xks}, BNS mergers may serve as an ideal source class for future joint neutrino-GW observations.
It presents us with an opportunity to study and explore -- what are the prospects of synergic GW and neutrino observations in the upcoming future, given the next generation of planned GW and neutrino detectors? How should we plan the joint searches across multimessenger channels to study these sources?

In Ref.~\cite{Mukhopadhyay:2023niv} we explored the above questions in the context of the planned next-generation of IceCube -- IceCube-Gen2~\cite{IceCube-Gen2:2020qha}, to search for high-energy neutrinos, using GW triggers from the next generation of GW detectors -- Cosmic Explorer (CE)~\cite{Reitze:2019iox} and Einstein Telescope (ET)~\cite{Maggiore:2019uih}. We studied the prospects by implementing \emph{triggered-stacking searches} for high-energy neutrinos. CE and ET have large distance horizons for BNS mergers $\sim$ a few Gpc. The large rate of BNS mergers at such distance scales and the uncertainty in the neutrino emission timescales from such sources, present a challenge for triggered-stacking searches, due to the unknown time window for the searches combined with many triggers. This leads to the obvious question: how does one collect \emph{meaningful triggers} in the era of the next generation of detectors? We proposed to alleviate this problem using the sky localization capabilities of the GW detectors, providing us with a distance limit for collecting meaningful triggers. In the current work, we extend our analysis to search for UHE neutrinos from BNS mergers at the next-generation of radio neutrino detectors -- the Giant Radio Detector for Neutrino Detection (GRAND)~\cite{GRAND:2018iaj} and IceCube-Gen2 Radio~\cite{IceCube-Gen2:2020qha}, and the currently operational Radio Neutrino Observatory in Greenland (RNO-G)~\cite{RNO-G:2020rmc}, using GW signals from BNS mergers in CE and ET as triggers.

In the current era, a dedicated triggered search pipeline called LLAMA (Low-Latency Algorithm for multimessenger Astrophysics)~\cite{Countryman:2019pqq} was developed to search for high-energy neutrinos from LIGO/Virgo candidates. This uses a fixed time-window of $\pm 500$ s and the overlap between the GW and neutrino localization areas to perform the analysis. Bayesian analysis based search techniques for common sources of GW and high-energy neutrinos were discussed in Ref.~\cite{Bartos:2018jco}. The IceCube collaboration also performed searches for high-energy neutrinos from BNS mergers~\cite{IceCube:2020xks} including implementing unbinned maximum likelihood and Bayesian analysis~\cite{IceCube:2022mma}. But no significant and strong evidence has been found. GW triggered searches for low energy neutrinos in the context of BNS mergers have also been explored in Ref.~\cite{Kyutoku:2017wnb,Lin:2019piz}. GW triggered searches have also been implemented to search for low-energy neutrinos from other sources like core-collapse supernovae~\cite{Mukhopadhyay:2021gox,Mukhopadhyay:2022qmo}. Besides implementing the sky-localization based distance cuts for triggers, the current work is also novel in this way that, for the first time we propose synergic search strategies for UHE neutrinos and investigate such prospects in the context of upcoming detectors.

The paper is organized as follows. We discuss the radio neutrino detectors relevant for UHE neutrino detection in Sec.~\ref{sec:nudets}. In Sec.~\ref{sec:formalism} we summarize the formalism for performing GW triggered UHE neutrino searches. We present our results and highlight the prospects for such searches in Sec.~\ref{sec:res}. We summarize in Sec.~\ref{sec:summary} and discuss the implications of our work in Sec.~\ref{sec:disc}.
\section{Ultrahigh-energy neutrino detectors}
\label{sec:nudets}
The low fluxes of UHE neutrinos require the development of cost effective techniques to achieve their detection. Several next-generation experiments have wagered on the observation of the radio signal emitted by cascades generated, either in ice or in air, by the daughter particles of neutrinos, once they have interacted in the ice or in the ground \cite{2022NatRP...4..697G,2022arXiv221013560C}. Radio antennas are indeed scalable, robust and usually cheaper than other more sophisticated detectors, enabling their extensive deployment in large numbers, over large areas. 

Here we will focus on two outstanding instruments planning to reach a diffuse source sensitivity of $\sim 2\times 10^{-10}\,{\rm GeV}\,{\rm cm}^{-2}\,{\rm sr}^{-1}\,{\rm s}^{-1}$ over 10 years, in the long-term future: GRAND~\cite{GRAND:2018iaj} and IceCube-Gen2 Radio \cite{IceCube-Gen2:2020qha}. Their timeline is consistent with the construction of next-generation GW instruments, opening the possibility of coincident neutrino-GW detections. 

In order to provide an insight of the up-coming possibilities at shorter timescales, we also examine the case of RNO-G~\cite{RNO-G:2020rmc}, which can be considered as a precursor of IceCube-Gen2 Radio in terms of experimental concept. This comparison will also highlight how the increase in sensitivity by more than an order of magnitude between RNO-G and GRAND/IceCube-Gen2 Radio can make a drastic difference on multimessenger detection prospects. 

\subsection{GRAND}
GRAND is a proposal for a set of $\sim 20$ radio antenna sub-arrays of 10\,000 km$^2$ each deployed on several continents, totaling 200\,000 antennas over 200\,000 km$^2$ by the 2030s \cite{GRAND:2018iaj,2023arXiv230800120G}. Its detection principle is as follows: Earth-skimming tau neutrinos interact in the ground, producing a tau particle, which in turn can emerge in the atmosphere and generate an extensive air-shower. The electromagnetic signal emitted by this shower can be observed by radio antenna arrays. By construction, the instantaneous field of view of each GRAND sub-array is hence reduced to $\sim 6^\circ$ below the horizon, for a $360^\circ$ azimuthal view, corresponding to $\sim 6\%$ of the sky per site. The total instantaneous field of view hence scales roughly with the number of sub-arrays. With sites located in both hemispheres, the day-averaged sky coverage reaches 100\% of the sky. In its ultimate configuration, the instrument will reach $10^{-1} {\rm GeV cm}^{-2}$ instantaneous point source sensitivity (assuming the zenith angle to be $90^\circ$) for neutrino energies $\sim 10^9$ GeV. Thanks to the kilometer-size radio footprints on the ground, an exquisite sub-degree angular resolution is expected to be reached~\cite{2023APh...14502779D}. 

\subsection{IceCube-Gen2 Radio and RNO-G}
IceCube-Gen2 Radio and RNO-G are based on the radio-detection of the coherent Askaryan emission initiated by any neutrino showers in ice. The in-ice technique offers better performances at lower energies ($\sim 10^{17}\,$eV, where the flux should be higher) than in-air techniques, which are strongly limited in that range by the tau exit probability from the ground. This enhancement comes however at the cost of a poorer angular resolution due to radio propagation effects in polar ice. The in-ice technique is also sensitive to all flavors of neutrinos, offering a complementary approach.

RNO-G is currently under construction. When completed, it will consist of 35 stations composed of 24 radio antennas, some installed at the surface of the ice and some buried underneath at various depths. 
The peak effective area will be the largest achieved to-date with a radio array, which is estimated to be $\gtrsim 10^{8} {\rm cm}^2$ for neutrino energies $\gtrsim 10^9$ GeV. The instantaneous field of view is mostly determined by the geometry of its location in Greenland, corresponding to roughly $45^\circ \times 360^\circ$. Due to the high-latitude of the site, the day-average leads to a coverage of half the Northern hemisphere sky. The preliminary estimates on the angular reconstruction points towards a neutrino arrival direction resolution at the few degree scale ($3^\circ \times 10^\circ$). 

IceCube-Gen2 Radio plans to build upon the experience of RNO-G. Located in the South Pole, it is planned to cover 500 km$^2$ with 361 stations~\cite{Gen2_TDR}. The peak effective area in this case, is projected as $\gtrsim 10^{10} {\rm cm}^2$ for neutrino energies $\gtrsim 10^{10}$ GeV. The instantaneous field of view is expected to be of $55^\circ \times 360^\circ$, corresponding to 43\% of the sky. At the pole, the rotation of the Earth does not affect the daily field of view. The angular resolution is expected to be similar to RNO-G, although improvements due to better analysis and reconstruction of the electric field polarization are expected. 

\section{Formalism}
\label{sec:formalism}
In this section we will briefly discuss the formalism for performing GW triggered searches for UHE neutrinos at the neutrino detectors using triggers from CE, ET, and CE+ET. A more rigorous description of the formalism can be found in Ref.~\cite{Mukhopadhyay:2023niv}. The prospects of UHE neutrino detection will depend on the properties of the source, the neutrino detector, and the relevant backgrounds. Thus the key ingredients required for quantifying the detection prospects are a typical profile for a flux of UHE neutrinos from BNS mergers, the redshift-dependent rate of BNS mergers, the energy and declination-dependent effective areas for the neutrino detectors, and background fluxes.

\subsection{UHE neutrino flux from BNS mergers}
\label{subsec:bnsnuflux}
BNS mergers, including short GRBs, can be accelerators of high-energy cosmic rays. The cosmic rays can interact with the surrounding matter and photons through hadronic interactions. In particular, the photomeson production ($p\gamma$) channel is relevant in jetted sources such as short GRBs. The resultant charged pions decay to muons which further decay to neutrinos with a typical energy of $1/20$ times the parent proton energy, although other mesons such as kaons and charmed mesons may become dominant at the highest energies~\cite{Carpio:2020wzg}. 

In this work, for simplicity, we focus on the generic power-law assumption without specifying neutrino production mechanisms. Then, the UHE neutrino fluence from a typical BNS merger source can be written as
\be
\label{eq:flux}
\phi_\nu (\mathcal{E}_\nu^{\rm UHE,iso}, E_\nu ,d_L) = \frac{(1+z)}{4 \pi d_L^2} \frac{\mathcal{E}_\nu^{\rm UHE,iso}}{\rm ln \big( \varepsilon_\nu^{\rm max}/\varepsilon_{\nu}^{\rm min} \big)} E_\nu^{-2}\,,
\ee
where we have assumed a power-law neutrino spectra with index $2$. The maximum and minimum energy in the source frame are given as $\varepsilon_\nu^{\rm min} = 10^3$ GeV and $\varepsilon_\nu^{\rm max} = 10^8$ GeV respectively. The luminosity distance to the source is defined as $d_L$ and the neutrino energy in observer frame is given by $E_\nu$ and gets a redshift correction $E_\nu = \varepsilon_\nu/(1+z)$, where $z$ is the redshift. The total UHE neutrino isotropic-equivalent energy is defined as $\mathcal{E}_\nu^{\rm UHE,iso}$
\begin{equation}
\label{eq:enuuheiso}
\mathcal{E}_\nu^{\rm UHE,iso} = \frac{\mathcal{E}_\nu^{\rm UHE,true}}{f_{\rm bm}}\,,
\end{equation}
where $\mathcal{E}_\nu^{\rm UHE,true}$ is the true energy emitted in UHE neutrinos at the source and $f_{\rm bm}$ is the beaming fraction. We assume $f_{\rm bm} \sim 1$\%, motivated from afterglow observations~\cite{Fong:2015oha}. The quantity $\mathcal{E}_\nu^{\rm UHE,true}$ is important and has a big influence on the detection prospects. This is intuitive since the more energy emitted in UHE neutrinos the better will be the detection prospects and lesser the detection timescales. Thus, we will use $\mathcal{E}_\nu^{\rm UHE,iso}$ as a parameter for this work. Neutrino searches will be sensitive to the isotropic-equivalent energy rather than the true energy. So whether the model has a degeneracy with the beaming factor is important and this can be solved with more detailed observations in principle.

\subsection{Triggered UHE neutrino searches}
We outlined a typical UHE neutrino fluence from a BNS merger in the previous section, which now puts us in a position to consider the UHE neutrino detection prospects and discuss a method to perform GW triggered neutrino searches. The total number of neutrinos $N_\nu$ in a neutrino detector given a flux $\phi_\nu (\mathcal{E}_\nu^{\rm UHE,iso}, E_\nu ,d_L)$ depends on the neutrino energy and the instantaneous zenith (or declination)-dependent effective area $\mathcal{A}_{\rm eff}(E_\nu, \theta_z)$
\be
\label{eq:nnu}
N_{\nu} (\theta_z,d_L) = \int_{E_{\nu}^{\rm LL}}^{E_{\nu}^{\rm UL}} dE_{\nu} \phi_\nu \big(\mathcal{E}_\nu^{\rm UHE,iso}, E_\nu ,d_L \big)\mathcal{A}_{\rm eff}(E_{\nu},\theta_z)\,,
\ee
where, the upper (lower) limits for neutrino energies are chosen to be $\
E_\nu^{\rm UL} = 10^{11}$ GeV $\big( E_\nu^{\rm LL} = 10^{7}$ GeV$\big)$. There are a few important things to note here. Although we fix the limits, GRAND becomes drastically less sensitive below energy $\lesssim 10^8$ GeV, due to the combined effects of a drop in tau (daughter particle of the tau neutrino) exit probability from the ground, and the radio signal becoming dominated by the Galactic noise in this regime.

For this work, we use the zenith-dependent effective area for GRAND~\cite{GRAND:2018iaj}, IceCube-Gen2 Radio~\cite{IceCube-Gen2:2021rkf,Gen2_TDR}, and RNO-G~\cite{RNO-G:2020rmc}. In particular, for IceCube-Gen2 Radio the simulations for effective area were produced using NuRadioMC~\cite{Glaser:2019cws}. The effective areas are incorporated in calculation of the number of events ($N_\nu$) from the above equation which then gives the Poissonian probability to detect at least one neutrino as, $p_{n\geq1} (\theta_z,d_L) = 1 - {\rm exp}\big( -N_\nu (\theta_z,d_L) \big)$. Integrating over the solid angle $\Omega$ leads to the total probability to detect at least one neutrino as
\be
\label{eq:totprob}
P_{n\geq 1} (d_L) = \frac{1}{N_{\rm norm}} \int d\Omega\ p_{n \geq 1} \big(\theta_z, d_L \big)\,,
\ee
where $N_{\rm norm}$ is the solid angle over which the probability is averaged. For all-sky, $N_{\rm norm} = 4\pi$.

Until now we have not considered the information that could be derived from the next-generation GW detectors. The BNS mergers detected by the GW detectors can be used as triggers to search for UHE neutrinos in the neutrino detectors. The probability to detect more than one neutrino associated with a GW signal can be given as
\begin{align}
\label{eq:qtop}
q\big( d^{\rm UL}_{\rm GW}, T_{\rm op} \big) &= 1 - {\rm exp}\bigg( -T_{\rm op} I \big( d_{\rm GW}^{\rm UL} \big) \bigg) \nonumber\,,\\
I \big( d_{\rm GW}^{\rm UL} \big) = 4\pi \int_{0}^{d_{\rm GW}^{\rm UL}} d (&d_{\rm com}) \frac{T_{\rm op}}{\big( 1+z \big)} R_{\rm app}\big( z \big) d_{\rm com}^2 P_{n \geq 1} \big( d_L \big)\,,
\end{align}
where $d^{\rm UL}_{\rm GW}$ is defined as the limiting distance for accepting triggers from the GW detectors to perform the neutrino searches, $R_{\rm app}(z)$ is the apparent redshift-dependent BNS merger rate, and $T_{\rm op}$ is the operation time of the GW detector. The integration is performed over the comoving distance $d_{\rm com}$ which is related to $d_L$ and $z$, $d_L = (1+z)d_{\rm com}$. For our analysis of triggered stacking searches, given the large $d_L$ for sources, $N_\nu << 1$. This, combined with accounting for the field of view (FOV) for the neutrino detectors the above equation can be written as (see Appendix~\ref{appsec:effarea} for details and derivation)
\begin{widetext}
\begin{equation}
\label{eq:maineq}
I \big( d_{\rm GW}^{\rm UL} \big) = 4\pi \int_{0}^{d_{\rm GW}^{\rm UL}} d (d_{\rm com}) \frac{T_{\rm op}}{\big( 1+z \big)} R_{\rm app}\big( z \big) d_{\rm com}^2
\left[ \int_{E_{\nu}^{\rm LL}}^{E_{\nu}^{\rm UL}} dE_{\nu} \phi_\nu \big(\mathcal{E}_\nu^{\rm UHE,iso}, E_\nu ,d_L \big)  \frac{1}{2} \int d({\rm cos}\theta_z) \mathcal{A}_{\rm eff}(\theta_z,E_\nu)  \right]\,.
\end{equation}
\end{widetext}

The redshift dependent rate of BNS mergers is derived from the rate of short gamma-ray bursts (GRBs)~\cite{Wanderman:2014eza}. This true rate involves two branches which split at $z \approx 0.9$ and is proportional to the fiducial rate at $z= 0,\ R_{\rm true} (z=0) = R_0$. The fiducial rate is poorly constrained from the current observations and can vary between $10\ {\rm Gpc}^{-3}{\rm yr}^{-1} - 1700\ {\rm Gpc}^{-3}{\rm yr}^{-1}$~\cite{LIGOScientific:2021usb,KAGRA:2021duu}. We mostly choose a conservative value of $R_0 = 300\ {\rm Gpc}^{-3}{\rm yr}^{-1}$ unless otherwise stated. The GW detectors provide the value of the true fiducial rate $R_0$. However the neutrino detectors are sensitive to the apparent redshift-dependent rate $R_{\rm app}(z)$ defined as
\be
\label{eq:rhoapparent}
R_{\rm app}(z) = f_{\rm bm} R_{\rm true}(z) = f_{\rm bm} R_0 \mathcal{R}(z) = R_{\rm app,0} \mathcal{R}(z)\,,
\ee
where the fiducial apparent rate $R_{\rm app,0} = f_{\rm bm} R_0$ and $\mathcal{R}(z)$ gives the redshift dependent rate~\cite{Wanderman:2014eza,Mukhopadhyay:2023niv}, such that $R_{\rm true}(z) = R_0 \mathcal{R}(z)$. We consistently take this into account in our computations. Furthermore, from Eqns.~\eqref{eq:flux},~\eqref{eq:enuuheiso}, and~\eqref{eq:rhoapparent}, we see that the product $R_0 \mathcal{E}^{\rm UHE,true}_\nu$ is independent of the beaming factor $f_{\rm bm}$, in the limit when $N_\nu << 1$. Thus, showing the results for $R_0 \mathcal{E}^{\rm UHE,true}_\nu$ is also a valid choice. However, for our current work we highlight the beaming corrections in both the rate and the energy in UHE neutrinos, so we chose the present scheme of showing our results.
\subsection{Distance limits from sky localization in GW detectors}
In this section, we will focus on the role of the GW detectors in the triggered search technique described above and how it quantifies $d^{\rm UL}_{\rm GW}$ in Eqs.~\eqref{eq:qtop} or~\eqref{eq:maineq}. The triggered search technique described above would be reasonable if the time of neutrino emission peak post the BNS merger was certain. However as discussed in Sec.~\ref{subsec:bnsnuflux}, this time window $\delta t$ is highly uncertain. Moreover, the next generation GW detectors are estimated to be sensitive to BNS mergers from very high redshifts. In particular, CE and ET would be sensitive to BNS mergers from $z \sim 10$ and $z\sim 3.5$ respectively~\cite{galaxies10040090}. Given the rate of the BNS mergers this implies $\mathcal{O}(100)$ BNS mergers detected in a day~\cite{galaxies10040090}. Given the large number of triggers, it then becomes important to have a selection criteria for \emph{meaningful triggers} and quantify the same, such that, the triggered search technique described above can still yield useful results.

There are multiple ways to address the above issue. For this work, we use the sky localization capabilities of the next-generation GW detectors to achieve the desired result. This in turn provides an upper limit on the distance from which meaningful triggers can be collected - $d^{\rm UL}_{\rm GW}$. Let us now briefly discuss the steps to evaluate $d^{\rm UL}_{\rm GW}$. Assume the GW detector detects a BNS merger at some distance $d_L$ and provides a trigger at some time $t_0$. The GW detector has a sky localization capability quantified by the size of the error or uncertainty region $\Delta \Omega (d_L)$ given a distance $d_L$. This depends on the properties of the detector. For example, ET has an equilateral configuration where the signal events can be triangulated, leading smaller $\Delta \Omega$ as compared to CE for the same distance $d_L$.

To search for a neutrino event corresponding to the trigger at $t_0$ one needs a time-window $\delta t$. However within this time $\delta t$, given the rate of BNS mergers $R_{\rm true}(z)$ there will be more signal detections by the GW detector each with a corresponding distance and hence a corresponding uncertainty region in the sky. Thus, at any point in time $\delta t$ post the trigger at $t_0$ the total fraction of sky area covered ($f_{\rm cov}$) due to the uncertainty regions of the subsequent detections by the GW detector, within the time-window $\delta t$ can be given as
\be
\label{eq:fth}
\int_0^{d_{\rm GW}^{\rm lim}} d\big( d_{\rm com} \big) \frac{\Delta \Omega \big(d_L \big)}{4 \pi} R_{\rm true}\big( z \big) 4 \pi d^2_{\rm com} \delta t = f_{\rm cov} (d_{\rm GW}^{\rm lim})\,.
\ee
The size of the uncertainty area $\Delta \Omega$ for CE, ET, and CE+ET is implemented from Ref.~\cite{Chan:2018csa}. Given the quality of data needed for the search one can choose a threshold value for the fraction of sky-area covered, say $f_{\rm th}$, that is, if at any given time post $t_0$, $f_{\rm cov} > f_{\rm th}$ we stop accepting any more triggers. This fixes the limiting distance for the GW detectors by demanding $d_{\rm GW}^{\rm lim}$ to be 
such that $f_{\rm cov} (d_{\rm GW}^{\rm lim}) = f_{\rm th}$. Thus the more signal events farther away hence low in SNR and have larger uncertainty area, the quicker $f_{\rm cov}$ reaches $f_{\rm th}$, the lesser the chances of such meaningless triggers to be accepted. Smaller values of $f_{\rm th}$ would be suitable for good quality data with lesser $d_{\rm GW}^{\rm lim}$, while larger values would be suited for constraining various neutrino emission models. Thus for this work, we will choose $f_{\rm th}$ depending on the properties of the detector.

We also fix a maximum and minimum distance horizon for the GW detectors as $d_{\rm GW}^{\rm hor,max}$ and $d_{\rm GW}^{\rm hor,min}$ respectively. The maximum distance horizon is chosen such that beyond this distance all triggers can be used. We fix this to the upper limit of the distance horizon for BNS mergers for ET, $z_{\rm GW}^{\rm hor,max} \sim 3.5$. This choice has no effect on our results since the BNS merger rate decreases for $z \gtrsim 0.9$. The minimum distance horizon is chosen to $d_{\rm GW}^{\rm hor,min} \approx 500$ Mpc. This choice is motivated by the maximum horizon distance of LIGO A\# (A-sharp)~\cite{T2200287}. This is reasonable because although we focus on the next-generation of GW detectors in this work, the timescales for realizations of ET and CE would be such that we will also have the improved next-generation of LIGO detectors. Such detectors can help us improve the localization for nearby sources. Besides, upcoming EM telescopes like the Vera C. Rubin Observatory's Legacy Survey of Space and Time (LSST)~\cite{LSST:2008ijt} or Roman~\cite{DES:2017dgt} can detect kilonovae as a result of BNS mergers from redshifts $z\sim 0.1$ and $z\sim 0.2$ respectively. This would greatly improve the sky localizations over just the GW observations. This justifies our choice of the minimum distance horizon. Finally, putting it altogether we have, 
\be
\label{eq:dulgw}
d^{\rm UL}_{\rm GW} = {\rm max}\left[ d_{\rm GW}^{\rm hor,min}, {\rm min} \left( d_{\rm GW}^{\rm lim}, d_{\rm GW}^{\rm hor,max} \right) \right]\,.
\ee
\section{Results}
\label{sec:res}
\begin{figure*}[ht!]
\includegraphics[width=0.98\textwidth]{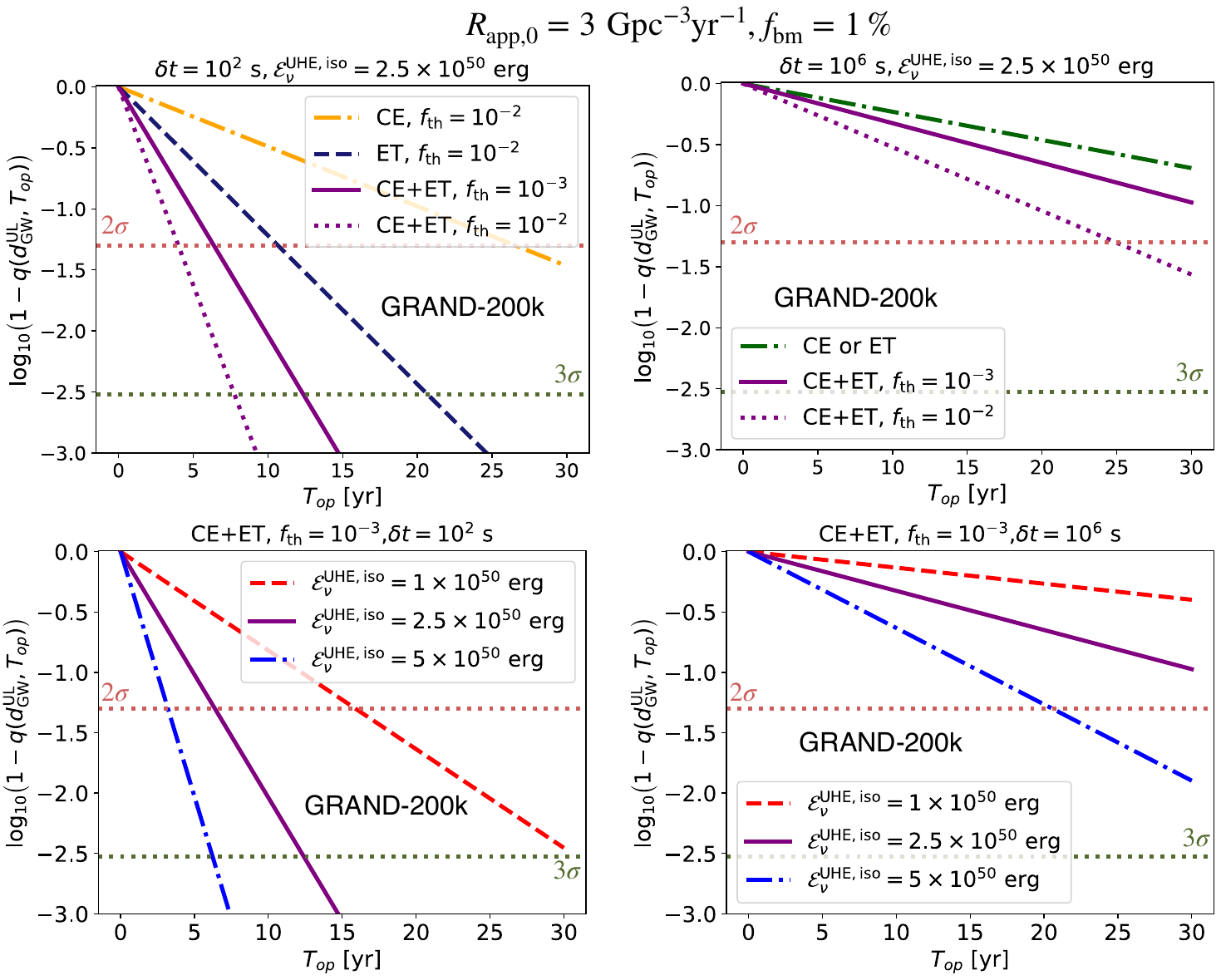}
\caption{\label{fig:grand_tops} 
Probability of coincident detection of GW-triggered UHE neutrinos $q \big( d_{\rm GW}^{\rm UL},T_{\rm op} \big)$ (hence $1-q \big( d_{\rm GW}^{\rm UL},T_{\rm op} \big)$ gives the probability to not see a coincident neutrino event) with the operation timescale of GW detectors $T_{\rm op}$ for GRAND-200k. The \emph{left} panels show the results when $\delta t = 10^2$ s (instantaneous) while the \emph{right} panels show the results when  $\delta t = 10^6$ s. The \emph{top} panels show the results from using different GW detectors whereas the \emph{bottom} panels highlight how the prospects depend on $\mathcal{E}_\nu^{\rm UHE,iso}$. The fiducial case for each panel is shown with a thick solid purple line, where we choose $\mathcal{E}_\nu^{\rm UHE,iso} = 2.5 \times 10^{50}$ erg. In case of nondetection, the $2\sigma$ and $3\sigma$ C.L.s with which the parameter space can be constrained are also shown with dotted horizontal lines. We choose the true fiducial rate $R_0 = 300\ {\rm Gpc}^{-3}{\rm yr}^{-1}$, however, the neutrino searches are sensitive to $R_{\rm app}(z)$, see Eq.~\ref{eq:rhoapparent}), for which we choose $f_{\rm bm} = 1$\% and hence $R_{\rm app,0} = 3\ {\rm Gpc}^{-3}{\rm yr}^{-1}$.
}
\end{figure*}
We discuss the main results of our work in this section. In particular, we distinguish between the upcoming UHE neutrino detectors -- GRAND and IceCube-Gen2 Radio, and the currently operating detector RNO-G. The results are presented with the aim to investigate the prospects in terms of timescale of operation of the GW detectors for synergic GW triggered neutrino observations or putting constraint at a given C.L. (confidence level) in case of nondetections.

The results are shown as plots of ${\rm log}_{10} \big(1-q( d^{\rm UL}_{\rm GW}, T_{\rm op} ) \big)$ (see Eq.~\ref{eq:qtop}) with the operation time of the GW detectors in years. Recall that $q\big( d^{\rm UL}_{\rm GW}, T_{\rm op} \big)$ gives the probability to detect one or more \emph{coincident} neutrino(s) associated with a GW signal\footnote{Note that here we only consider the probability of neutrino detections associated with the GW trigger. So it is equivalent to saying that the neutrino detection significance is $\sim 5\sigma$ and we will consider this to be the case for the remaining text when we describe \emph{coincident neutrino detection}. However, the association of the event with BNS mergers is non-trivial. This requires a separate analysis pertaining to the probability of random associations between the neutrino event and the BNS mergers. A $5\sigma$ C.L. coincident neutrino detection from a BNS merger can only be claimed if \emph{both} the neutrino detection and association significance are above $5\sigma$. For association significance, the error region associated with the neutrino events will also play an important role.}, thus when $q\big( d^{\rm UL}_{\rm GW}, T_{\rm op} \big)$ is 0, we have ${\rm log}_{10} \big(1-q( d^{\rm UL}_{\rm GW}, T_{\rm op} ) \big)$ to be 0 corresponding to 0 years of operation. As the operation time for the GW detectors increase, $q\big( d^{\rm UL}_{\rm GW}, T_{\rm op} \big)$ increases and ${\rm log}_{10} \big(1-q( d^{\rm UL}_{\rm GW}, T_{\rm op} ) \big)$ decreases. Thus lower values of ${\rm log}_{10} \big(1-q( d^{\rm UL}_{\rm GW}, T_{\rm op} ) \big)$ indicate higher probabilities of finding a neutrino event corresponding to GW triggers. We also show the $95$\% ($2 \sigma$) and $99.7$\% ($3 \sigma$) C.L.s to constrain the associated parameter space in case of a non-detection by dotted dark red and dark green horizontal lines respectively.

We fix the true fiducial rate $R_0 = 300\ {\rm Gpc}^{-3}{\rm yr}^{-1}$ and choose $f_{\rm bm} = 1$\%. This gives the fiducial apparent rate as $R_{\rm app,0} = 3\ {\rm Gpc}^{-3}{\rm yr}^{-1}$ (see Eq.~\ref{eq:rhoapparent}). The next parameter that we need to choose to perform our analysis is the threshold fraction of sky area covered $f_{\rm th}$. This, in general, is motivated by the extent to which the detectors can perform sky localization and is limited by the background in the neutrino detector associated with triggered searches. At the neutrino energies of interest for this work, the atmospheric background is minimal and the astrophysical background is estimated to be low\footnote{Note that here astrophysical background refers to the diffuse astrophysical background measured by IceCube~\cite{Naab:2023xcz}.}. 
Thus it is possible to choose relatively large values of $f_{\rm th}$. In principle, we can choose the most optimistic value of $f_{\rm th} \sim 10^{-2}$ for all the detectors. However, to give conservative estimates we choose $f_{\rm th} = 10^{-2}, 10^{-2},$ and $10^{-3}$ for CE, ET, and the combination of CE+ET. Finally, the two main parameters varied to compute the results are the isotropic-equivalent energy emitted in UHE neutrinos ($\mathcal{E}_\nu^{\rm UHE,iso}$) and the time duration post the merger ($\delta t$) when the neutrino emission peak is reached, which is the same as the size of the search window. 

\begin{figure*}[ht!]
\includegraphics[width=0.98\textwidth]{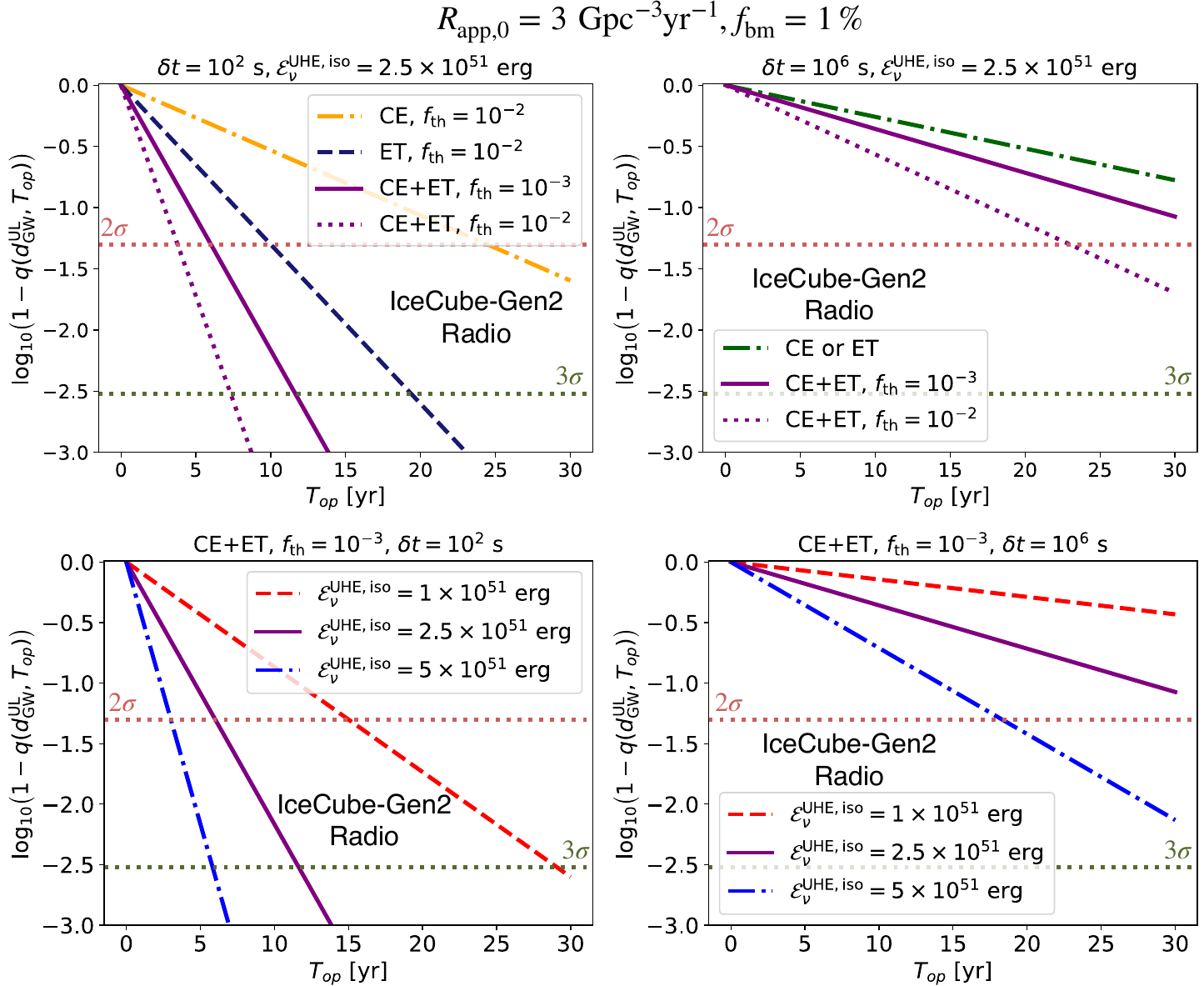}
\caption{\label{fig:icg2rad_tops} Same as Fig.~\ref{fig:grand_tops} but for IceCube-Gen2 Radio. For IceCube-Gen2 Radio we choose the fiducial case as $\mathcal{E}_\nu^{\rm UHE,iso} = 2.5 \times 10^{51}$ erg.
}
\end{figure*}
\begin{figure*}[ht!]
\includegraphics[width=0.98\textwidth]{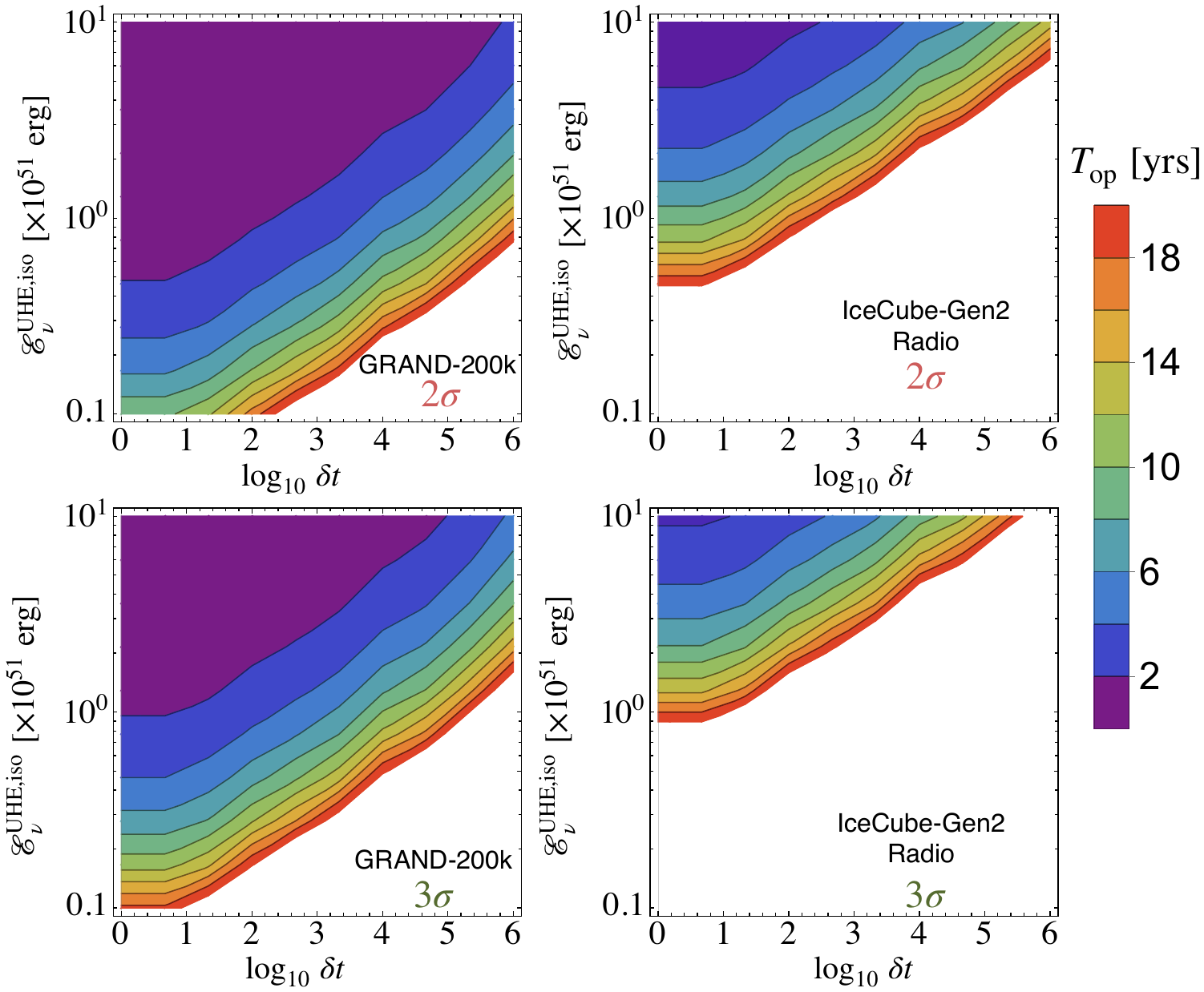}
\caption{\label{fig:cl_en_delt} Contour plots highlighting the joint GW and neutrino detector operation timescales ($T_{\rm op}$) in years for GW-triggered UHE neutrino searches. The contours in the \emph{upper} panel denote $T_{\rm op}$ for the GW and neutrino detectors required to have a $95$\% probability to detect coincident UHE neutrino events or constrain the parameter space at the $2\sigma$ level in case of nondetections for GRAND (\emph{left} panels) and IceCube-Gen2 Radio (\emph{right} panels). The \emph{lower} panels also show the same contours for the same detectors, but for $99.7$\% probability to detect coincident UHE neutrino events or constrain the parameter space at the $3\sigma$ level in case of nondetections. The search is performed assuming a combination of CE+ET as the GW detectors with $f_{\rm th} = 10^{-3}$ and the true fiducial rate $R_0 = 300\ {\rm Gpc}^{-3}{\rm yr}^{-1}$ (this is the true rate, however the neutrino searches are sensitive to $R_{\rm app}(z)$, see Eq.~\ref{eq:rhoapparent}). We choose $f_{\rm bm} = 1$\%.
}
\end{figure*}

For the fiducial values of $\delta t$, we divide it into two classes -- instantaneous emission where $\delta t = 10^2$ s and the case when the neutrinos are emitted $\mathcal{O}(10)$ days post the merger, that is, $\delta t = 10^6$ s. The instantaneous case yields more optimistic results than the case when $\delta t = 10^6$ s. This is because as $\delta t$ increases, the number of GW triggers within $\delta t$ also increases. This leads to a decrease in the limiting distance $d_{\rm GW}^{\rm lim}$ for a fixed value of $f_{\rm th}$. Thus the upper limit of in the integral in Eqs.~\eqref{eq:qtop} or~\eqref{eq:maineq} goes down leading to the requirement of longer operation timescales. We give the details of various parameters in Table~\ref{tab:params}.

\begin{table}
\centering
\begin{tabular}{|>{\centering\arraybackslash}p{0.03\textwidth}|>{\centering\arraybackslash}p{0.1\textwidth}|>{\centering\arraybackslash}p{0.05\textwidth}|>{\centering\arraybackslash}p{0.1\textwidth}|>{\centering\arraybackslash}p{0.1\textwidth}|}
\hline
 $\delta t$ &  GW &  $f_{\rm th}$ &  $d_{\rm GW}^{\rm UL} [z]$ & $N_{\rm trig}$ \\
(s) & Detectors & & (Gpc) & (yr$^{-1}$)\\
\hline
\hline
     & CE    & $10^{-2}$ & $1.07\ [0.21]$ & $9.6 \times 10^{1}$\\
 100 & ET    & $10^{-2}$ & $2.80\ [0.48]$ & $1.6 \times 10^{3}$\\
     & CE+ET & $10^{-3}$ & $4.72\ [0.74]$ & $7.7 \times 10^{3}$\\
     & CE+ET & $10^{-2}$ & $8.47\ [1.20]$ & $3.3 \times 10^{4}$\\
\hline
\hline
       & CE or ET & -   & $0.50\ [0.10]$ & $9.1 \times 10^{0}$\\
$10^6$ & CE+ET    & $10^{-3}$ & $0.70\ [0.14]$ & $2.6 \times 10^{1}$\\
       & CE+ET    & $10^{-2}$ & $1.14\ [0.22]$ & $1.1 \times 10^{2}$\\
\hline
\end{tabular}
\caption{\label{tab:params} Parameters relevant for computing $d_{\rm GW}^{\rm UL}$ (see Eq.~\ref{eq:dulgw}) for different GW detectors for the instantaneous $\delta t = 100$ s and $\delta t = 10^6$ s scenarios. The true fiducial rate is chosen to be $R_0 = 300\ {\rm Gpc}^{-3}{\rm yr}^{-1}$. The values of $d_{\rm GW}^{\rm UL}$ shown in the table are used to compute the corresponding $q( d^{\rm UL}_{\rm GW}, T_{\rm op} )$ (see Eq.~\ref{eq:qtop}) shown in all the figures. The number of GW triggered events per year ($N_{\rm trig}$) for a corresponding choice of detector and $f_{\rm th}$ is also shown. A fraction $f_{\rm bm} = 1$\% of $N_{\rm trig}$ are expected to contribute to on-axis sGRBs.}
\end{table}
\begin{figure*}[ht!]
\includegraphics[width=0.98\textwidth]{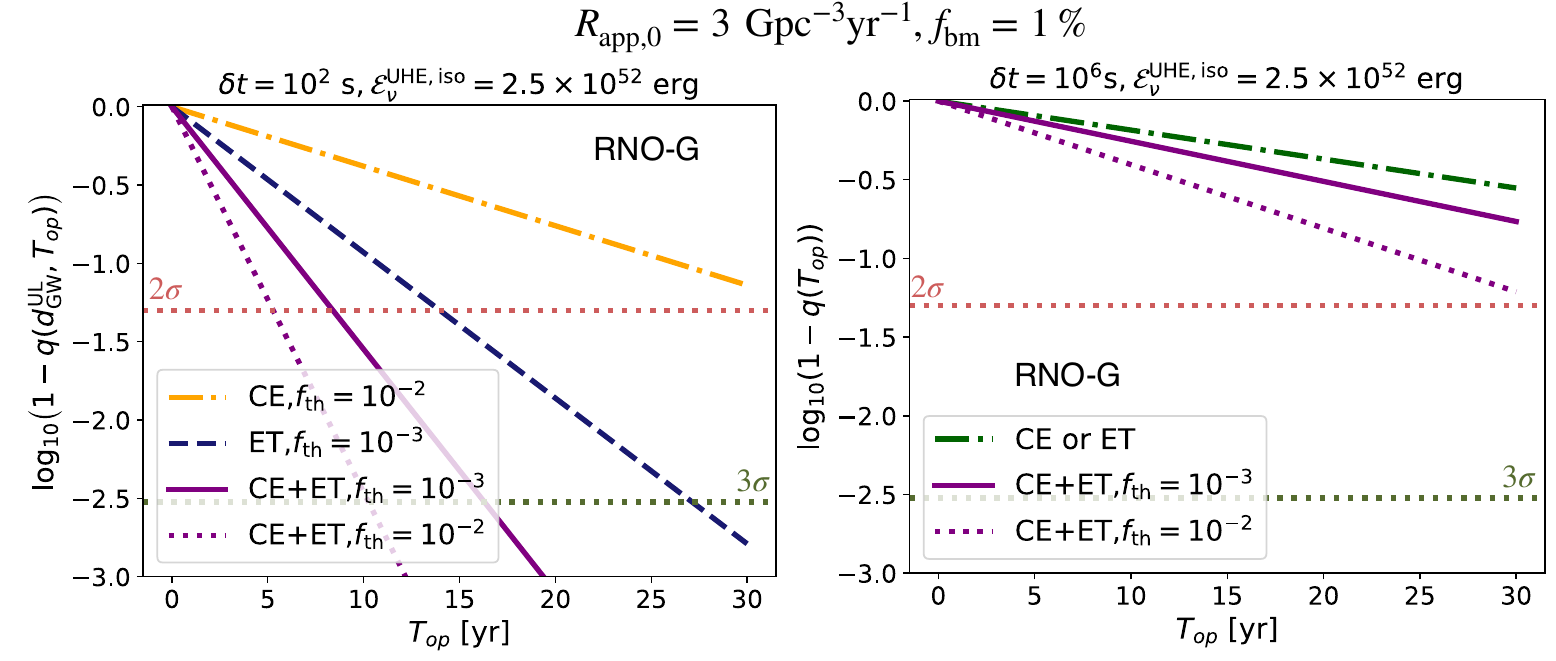}
\caption{\label{fig:rnog_tops} Same as Fig.~\ref{fig:grand_tops} but for RNO-G. For RNO-G we choose the fiducial case as $\mathcal{E}_\nu^{\rm UHE,iso} = 2.5 \times 10^{52}$ erg. We only focus on the results of using different GW detectors as triggers for the searches.
}
\end{figure*}
\begin{figure}
\includegraphics[width=0.49\textwidth]{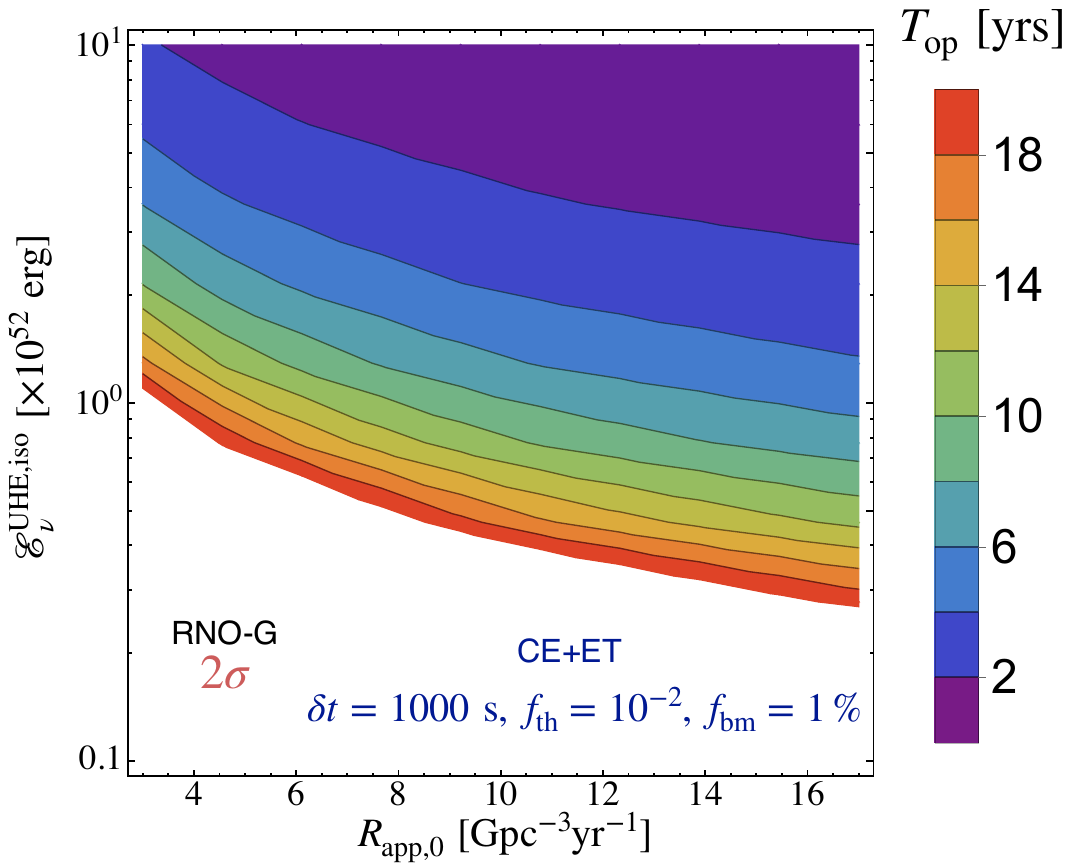}
\caption{\label{fig:rnog_rate_2sig} Same as Fig.~\ref{fig:cl_en_delt} but for RNO-G, where the operation timescales are shown in the $\mathcal{E}_\nu^{\rm UHE,iso}-R_{\rm app,0}$ plane. For the apparent rate we choose $f_{\rm bm} = 1$\% (see Eq.~\ref{eq:rhoapparent}). The search is performed assuming a combination of CE+ET as the GW detectors with $f_{\rm th} = 10^{-2}$ and $\delta t = 1000$ s. 
}
\end{figure}
\subsection{Prospects for upcoming UHE neutrino detectors: GRAND and IceCube-Gen2 Radio}
We show the results for GRAND and IceCube-Gen2 Radio in Figs.~\ref{fig:grand_tops} and~\ref{fig:icg2rad_tops} respectively. We choose the fiducial values of $\mathcal{E}_\nu^{\rm UHE,iso} = 2.5 \times 10^{50}$ erg and $2.5 \times 10^{51}$ erg for GRAND and IceCube-Gen2 Radio respectively. This is within reasonable limits based on predictions from various models of UHE neutrino emission from BNS mergers~\cite{Fang:2017tla,Kimura:2018vvz,Mukho_nu}. The lower values of instantaneous effective areas for IceCube-Gen2 Radio as compared to GRAND requires us to choose $\mathcal{E}_\nu^{\rm UHE,iso}$ one order of magnitude higher for the former (see Sec.~\ref{sec:formalism}). The \emph{top right} panels of Figs.~\ref{fig:grand_tops} and~\ref{fig:icg2rad_tops} show the results for the instantaneous fiducial case. Both for GRAND and IceCube-Gen2 Radio, assuming the fiducial case, there is $\sim 99$\% probability to detect coincident neutrino event/events, using triggers from CE+ET in a timescale of $\lesssim 15$ years. In case of a non-detection $3\sigma$ level constraints can be put on the parameter space in less than 15 years. Considering the most optimistic value of $f_{\rm th} = 10^{-2}$, increases the probability to $\sim 99$\% for coincident detections and decreases the time required to place $3\sigma$ level constraints in case of non-detections to $\lesssim$ 10 years. Using ET and CE individually, $3\sigma$ and $2\sigma$ level constraints can be put in timescales $\lesssim$ 25 and 30 years respectively, in the case of non-detection of UHE neutrinos. These results are optimistic and present reasonable chances of coincident neutrino detection or the opportunity to place constraints provided the neutrino emission is prompt post-merger. However it is important to keep in mind that although the timescales required for detection or placing constraints are similar for GRAND and IceCube-Gen2 Radio, our analysis based on the currently available effective areas, requires larger values (by a factor of 10) of energy emitted in UHE neutrinos for IceCube-Gen2 Radio to place the same constraints.

The \emph{top left} panels in Figs.~\ref{fig:grand_tops} and~\ref{fig:icg2rad_tops} show the fiducial case but for $\delta t = 10^6$ s. As expected for this case the results are less optimistic. We see that both for GRAND and IceCube-Gen2 Radio, the $90$\% probability to detect one or more coincident neutrino events requires a very long period of $\sim 20$ years of joint operation time even after assuming the most optimistic threshold $f_{\rm th} = 10^{-2}$ for CE+ET. Only $2\sigma$ level constraints in case of nondetections are possible in a joint operation timescale of $\sim 25$ years using CE+ET. For the individual detectors we find $d_{\rm GW}^{\rm UL}$ to be fixed at $d_{\rm GW}^{\rm hor, min} = 500$ Mpc (see Eq.~\ref{eq:dulgw}). For this case we can only have $\sim 80$\% chance of coincident GW triggered UHE neutrino detections.

The \emph{bottom left} and \emph{bottom right} panels of Fig.~\ref{fig:grand_tops} (GRAND) and Fig.~\ref{fig:icg2rad_tops} (IceCube-Gen2 Radio) show the effects of varying the total energy emitted in UHE neutrinos ($\mathcal{E}_\nu^{\rm UHE,true}$) for the instantaneous case ($\delta t = 100$ s) and $\delta t = 10^6$ s respectively. In these panels we only show the results assuming CE+ET working together and choosing the fiducial value of $f_{\rm th} = 10^{-3}$. As a general trend we see that the more the energy emitted in UHE neutrinos, the greater the number of events and lesser the timescales of operation to have coincident neutrino detections or stronger constraints in the case of nondetections. We vary $\mathcal{E}_\nu^{\rm UHE,iso}$ between $10^{50}$ erg to $5 \times 10^{50}$ erg $10^{51}$ erg to $5 \times 10^{51}$ erg for GRAND and IceCube-Gen2 Radio respectively. 

For the upper limit of $\mathcal{E}_\nu^{\rm UHE,iso}$ we have the best case scenario for the instantaneous case, where there is a $\sim 99$\% probability to detect a coincident neutrino event in $\sim 7$ years of joint operation of the neutrino and GW detectors. The $3\sigma$ level constraints in case of non-detection can be placed over similar timescales. Even for the least optimistic scenario individually for GRAND and IceCube-Gen2 Radio, the probability of coincident UHE neutrino detection is $\sim 99$\% in $30$ years or non-detection of the UHE neutrinos would lead to $3\sigma$ constraints on the parameter space over similar timescales of operation. When $\delta t =10^6$ s, it would take $\sim 20$ years of joint operation of the GW and the neutrino detectors to have a $95$\% probability to detect an event or place a $2\sigma$ level constraint in the case of non-detection. If $\mathcal{E}_\nu^{\rm UHE,iso}$ is close to the lower limit we predict the instantaneous scenario might still be able to have coincident neutrino observation in $\sim 15 - 20$ years with a probability $\sim 95$\%. However the prospects are not very promising if $\delta t = 10^6$ s, where the probability of such an observation is $<70$\%. This is to be expected, since this is a representation of a pessimistic scenario where $\mathcal{E}_\nu^{\rm UHE,iso}$ is low and $\delta t$ is large.

One of the main takeaways from this work is shown in Fig.~\ref{fig:cl_en_delt} where we investigate the prospects of coincident GW-triggered (using CE+ET with $f_{\rm th} = 10^{-3}$) UHE neutrino searches in the plane of $\mathcal{E}_\nu^{\rm UHE,iso}$ and $\delta t$ -- the two primary parameters in our analysis. The contours in the \emph{upper} panels denote the joint operation timescale $T_{\rm op}$ for the GW and neutrino detectors in years required to have a $95$\% probability to detect coincident UHE neutrino events or constrain the parameter space at the $2\sigma$ level in case of nondetections for GRAND (\emph{left} panel) and IceCube-Gen2 Radio (\emph{right} panel). The \emph{lower} panels also show the same contours for the same detectors, but for $99.7$\% probability to detect coincident UHE neutrino events or constrain the parameter space at the $3\sigma$ level in case of nondetections. We limit the timescale to a reasonable value of $20$ years of operation, so any part of the parameter space that would need more than $20$ years is shown in white. 

The most optimistic parameters lie in the top left corner where $\mathcal{E}_\nu^{\rm UHE,iso} \gtrsim 10^{51}$ erg and $\delta t \lesssim 100$ s. Note that for high-energy neutrinos this parameter space is ruled out based on sGRB limits from IceCube~\cite{IceCube-Gen2:2021rkf}. The current limits from high energy neutrinos are computed by considering $90$\% C.L. limits, obtained from the time-integrated flux at 1 TeV for all sGRBs assuming an injected spectrum of $E_\nu^{-2.28}$. Considering an average sGRB redshift distance of $z_{\rm GRB}^{\rm avg} = 0.5$, one can convert the flux into $\mathcal{E}_\nu^{\rm HE,iso}$. There are however some key differences in showing the same limits for $\mathcal{E}_\nu^{\rm UHE,iso}$ based on this analysis. Firstly, our study assumes a $E_\nu^{-2}$ spectrum which is harder than what was considered in Ref.~\cite{IceCube-Gen2:2021rkf}. Second, the energy ranges for high-energy neutrinos would be $10^{3}$ GeV to  $10^{6}$ GeV which is different from the UHE case, where the energy range is given by $10^{7}$ GeV to  $10^{11}$ GeV. 

Besides, the number of sGRBs obtained from implementing our current analysis would be different from the one obtained in Ref.~\cite{IceCube-Gen2:2021rkf}. This is because we use the GW triggers for calculating the number of sGRBs whereas, Ref.~\cite{IceCube-Gen2:2021rkf} used the number of sGRBs observed in the electromagnetic band. The number of sGRBs considered for the analysis in Ref.~\cite{IceCube-Gen2:2021rkf} was $317$ in 10 years, that is, $\sim 31$ per year. Let us look at some estimates related to the number of triggers that we would obtain from our analysis. Assuming a distance horizon for BNS mergers of $z \sim 3.2$ for the Einstein Telescope, gives $\sim 4.2 \times 10^4$ triggers per year. Note that this is without implementing the distance limits from considering the sky localization of the events. A fraction $f_{\rm bm} = 1$\% of these triggers are expected to contribute to on-axis sGRBs, that is, the number of such events are $\sim 4.2 \times 10^2$ per year, which is roughly an order of magnitude larger than what was used in Ref.~\cite{IceCube-Gen2:2021rkf}. On implementing the distance cuts, the number of events based on this will be even lower due to the reduced distance horizon. We show the number of GW-triggered events per year ($N_{\rm trig}$) given a detector and a choice of $f_{\rm th}$ in Table~\ref{tab:params} to provide an estimate.
\subsection{Current prospects: RNO-G}
The results for the currently operating RNO-G are shown in Fig.~\ref{fig:rnog_tops}. We show the prospects for the instantaneous case and $\delta t = 10^6$ s in the \emph{left} and the \emph{right} panels respectively. As discussed in Sec.~\ref{sec:nudets}, RNO-G is aimed to be a precursor to IceCube-Gen2 Radio and hence has comparatively smaller effective areas. Due to this we choose the fiducial value of $\mathcal{E}_\nu^{\rm UHE,iso}$ an order of magnitude higher than what we chose for IceCube-Gen2 Radio, that is, $\mathcal{E}_\nu^{\rm UHE,iso} = 2.5 \times 10^{52}$ erg. This of course, is a very optimistic choice and is unlikely. However, the goal of the results in this section is just to provide an estimate of the performance of RNO-G for the purpose of GW-triggered UHE neutrino searches.

For the most optimistic choice of $f_{\rm th} = 10^{-2}$ using CE+ET, RNO-G has a $99.7$\% probability to detect a coincident UHE neutrino event or constrain the parameter space at $3\sigma$ in the case of non-detection in a timescale of $\sim 10$ years for the instantaneous case when $\delta t = 100$ s. If $\delta t \sim 10^6$ s, the joint operation timescales required to have a $95$\% probability to detect a coincident UHE neutrino event is $\gtrsim 30$ years, which is similar to the timescale required to place constraints at the $2\sigma$ level in case of non-detections.

It is evident that given the fiducial estimates, RNO-G due its small size requires impractical timescales of joint operation to detect coincident neutrino events or constrain the parameter spaces. However recall that the fiducial true rate of BNS mergers have a huge uncertainty window ranging from $10\ {\rm GeV}^{-3}{\rm yr}^{-1}$ to $1700\ {\rm GeV}^{-3}{\rm yr}^{-1}$. Thus one can speculate assuming the energy emitted in UHE neutrinos is indeed optimistic, what sort constraints can RNO-G place in the plane of isotropic-equivalent energy emitted in UHE neutrinos ($\mathcal{E}_\nu^{\rm UHE,iso}$) and the apparent fiducial rate ($R_{\rm app,0}$)? This is what we show in Fig.~\ref{fig:rnog_rate_2sig}. We show the results in the apparent rate and the isotropic-equivalent energy emitted in UHE neutrinos plane, assuming the beaming factor to be $f_{\rm bm}=1$\%\footnote{ Note that the beaming factor important for neutrino production is not well-constrained and thus the relations between the isotropic-equivalent and the true energy and the true and the apparent rate could be non-trivial.}. We also assume that the neutrinos are emitted at a typical value of $\delta t\lesssim 1000$ s post-merger hence fixing the size of the time-window for the triggered search. The results manifest that given the assumptions, RNO-G might be able to put constraints on the apparent rates based on non-detections at the $2\sigma$ level in a joint operation timescale of $\gtrsim 8$ years. In the meantime, if more precise measurements of the true rate if BNS mergers are reported by the GW detectors, $\mathcal{E}_\nu^{\rm UHE,iso}$ can be constrained.

\section{Summary}
\label{sec:summary}
\begin{figure*}[ht!]
\centering
\subfloat[(a)] {\includegraphics[width=0.49\textwidth]{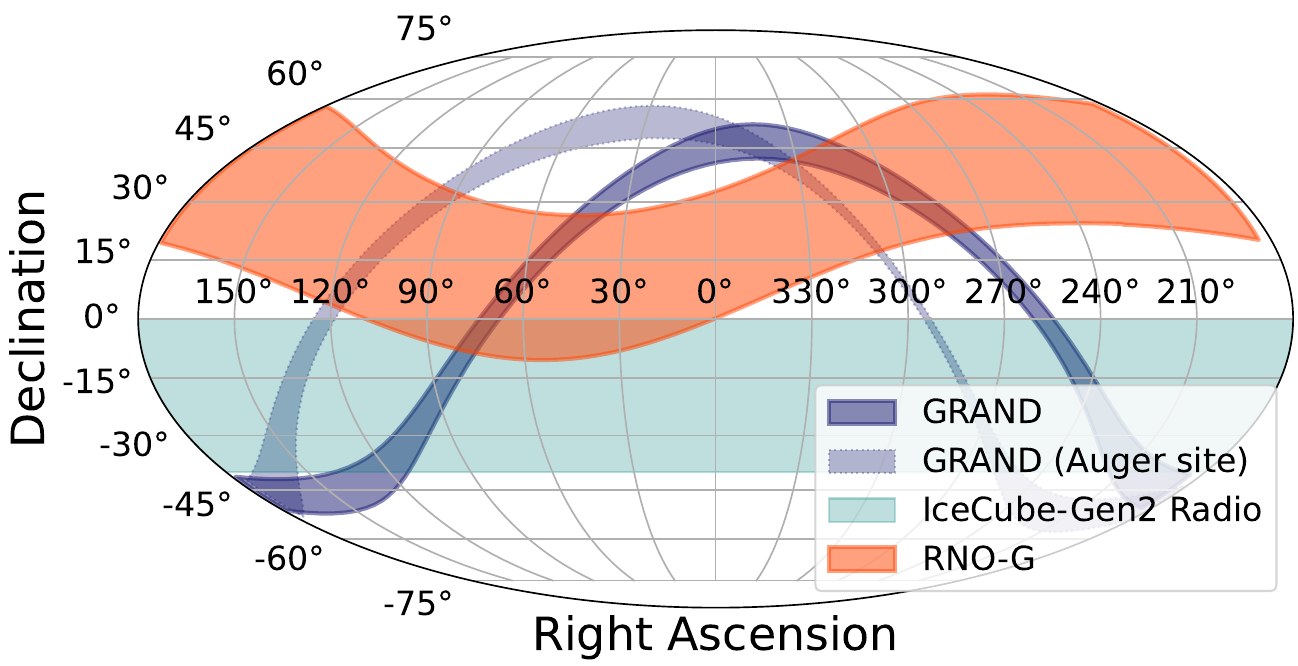}}\hfill
\subfloat[(b)] {\includegraphics[width=0.49\textwidth]{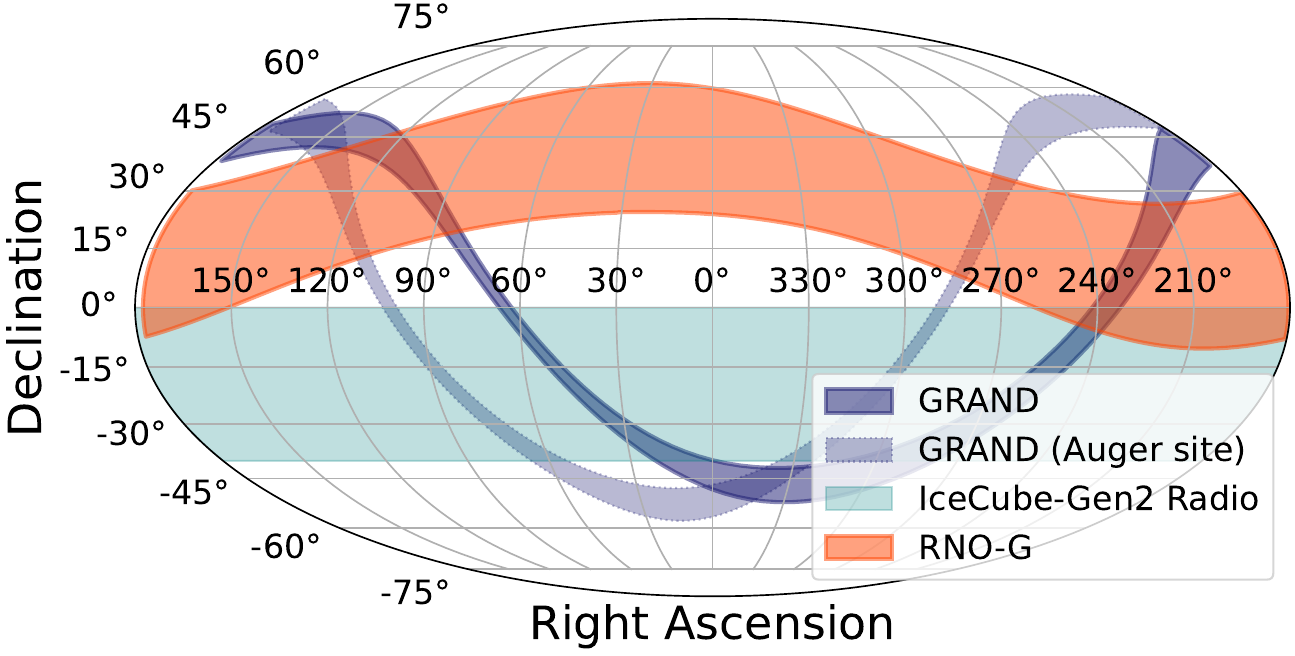}} \hfill
\caption{\label{fig:fov_nudets}The instantaneous field of view (FOV) for GRAND, IceCube-Gen2 Radio, and RNO-G for an arbitrarily chosen date of December 15, 2032 and time of the day 12:00:00 (\emph{left} panel) and 23:00:00 (\emph{right} panel) (this choice is arbitrary but reflects the complementary nature of the radio neutrino detectors). For IceCube-Gen2 Radio we only show the FOV which has the highest effective areas for neutrino energies $E_\nu \gtrsim 10^7$ GeV. The locations used for summit stations of the detectors are provided in Table~\ref{tab:nu_loc}. 
}
\end{figure*}
\begin{table}
\centering
\begin{tabular}{|>{\centering\arraybackslash}p{0.08\textwidth}|>{\centering\arraybackslash}p{0.1\textwidth}|>{\centering\arraybackslash}p{0.1\textwidth}|>{\centering\arraybackslash}p{0.07\textwidth}|>{\centering\arraybackslash}p{0.09\textwidth}|}
\hline
Neutrino Detectors & Latitude  &  Longitude & Elevation  (m)  &  Zenith Angle  \\
\hline
\hline
GRAND &  $43^\circ 00^\prime 00.00^{\prime \prime}$ N &  $80^\circ 00^\prime 00.00^{\prime \prime}$ E & 1300 & $85^\circ - 95^\circ$ \\
\hline
GRAND (Auger site) &  $37^\circ 12^\prime 48.62^{\prime \prime}$ S &  $55^\circ 06^\prime 35.10^{\prime \prime}$ W & 1000 & $85^\circ - 95^\circ$ \\
\hline
IceCube-Gen2 Radio & $90^\circ 00^\prime 00.00^{\prime \prime}$ S & $45^\circ 00^\prime 00.00^{\prime \prime}$ E & 2835 & $50^\circ - 90^\circ$ \\
\hline
RNO-G & $72^\circ 34^\prime 27.89^{\prime \prime}$ N & $38^\circ 27^\prime 19.83^{\prime \prime}$ W & 3216 & $46^\circ - 83^\circ$ \\
\hline
\end{tabular}
\caption{\label{tab:nu_loc} Location of the various radio neutrino detectors along with their zenith angle coverage, used to show the instantaneous FOV in Fig.~\ref{fig:fov_nudets}.
}
\end{table}
The search for ultrahigh-energy neutrinos from BNS mergers is timely because of the upcoming radio neutrino detectors. To maximize our chances of detecting UHE neutrinos from BNS mergers the GW-triggers associated with the latter can be utilized. In this work, we explore the prospects of GW-triggered stacking searches of UHE neutrinos at the radio neutrino detectors using GW triggers from the next-generation GW detectors like Cosmic Explorer (CE), Einstein Telescope (ET) and the combination of the two (CE+ET).

The probability of detect one or more coincident UHE neutrino event is given by Eq.~\eqref{eq:maineq}. This depends on the joint operation time of the GW and neutrino detectors, the redshift dependent rate of BNS mergers, the flux of UHE neutrinos, the zenith- and energy-dependent effective areas of the neutrino detectors, and their respective FOVs. For our work, we use the consistent zenith- and energy-dependent effective areas along with FOVs to compute our results. Furthermore, for the next generation of GW detectors, given their large distance horizon for BNS mergers, it is also important to collect \emph{meaningful} triggers to perform the triggered-stacking searches we present in this work. To achieve this, we use the sky-localization capabilities of the GW detectors and impose a maximum threshold of sky area covered ($f_{\rm th}$) to get an upper limit on the maximum distance ($d_{\rm GW}^{\rm UL}$) from which meaningful triggers can be collected (see Eqs.~\ref{eq:qtop} or~\ref{eq:maineq}).

We present our results in the form of probability to detect coincident UHE neutrinos ($q (d_{\rm GW}^{\rm UL}, T_{\rm op})$) with the timescale of joint operation for the GW and neutrino detectors ($T_{\rm op}$) in years. We also show in case of non-detections the time it would take to put $2\sigma$ and $3\sigma$ level constraints in the parameter space. We analyze two different scenarios of UHE neutrino emission timescales motivated by different physical models - an instantaneous emission with $\delta t = 100$ s and when the emission happens $\delta t = 10^6$ s post-merger. The former presents us with the most optimistic scenarios while the latter leads to far less optimistic results. 

In Fig.~\ref{fig:grand_tops} we show the results for GRAND. For the fiducial case, is UHE neutrino emission is instantaneous, assuming a very reasonable energy emitted in UHE neutrinos $\mathcal{E}_\nu^{\rm UHE,iso} = 2.5 \times 10^{50}$ erg, we find that the probability to see coincident UHE neutrino events is $99.7$\% in less than 15 years of joint operation with CE+ET detectors. Constraints at the $3\sigma$ level can be imposed in the case of non-detections in similar timescales using CE+ET. For $\delta t = 10^6$ s however, to have a $95$\% probability for a coincident detection or to place constraints at the $2\sigma$ level in the case of non-detection, roughly $25$ years oj joint operation is required, assuming CE+ET working with the most optimistic $f_{\rm th} = 10^{-2}$. For IceCube-Gen2 Radio, we show our results in Fig.~\ref{fig:icg2rad_tops}. The exact same conclusions can be arrived at by choosing the fiducial energy emitted in UHE neutrinos to be one order of magnitude higher, that is, $\mathcal{E}_\nu^{\rm UHE,iso} = 2.5 \times 10^{51}$ erg. This is a result of the instantaneous effective areas of GRAND being higher than that of IceCube-Gen2 Radio. We also show how the conclusions change if $\mathcal{E}_\nu^{\rm UHE,iso}$ is varied across a range. 

Probably our most important takeaway from this work is shown in Fig.~\ref{fig:cl_en_delt}. We present the prospects of joint GW-triggered UHE neutrino searches using CE+ET and choosing $f_{\rm th} = 10^{-3}$. We plot the contours for the joint operation timescales ($T_{\rm op}$) required to have a $95$\% and $99$\% probabilities to detect coincident UHE neutrino events or constrain the parameter space at the $2\sigma$ and $3\sigma$ levels in case of nondetections for GRAND and IceCube-Gen2 Radio. We note that GRAND, for $\mathcal{E}_\nu^{\rm UHE,iso} \gtrsim 2 \times 10^{51}$ erg, can probe the parameter space across all values of $\delta t$ for the desired probabilities and level of constraints in an operation timescale of $\sim 20$ years.

We also investigate the prospects for the current generation detector RNO-G. Since RNO-G is operating as a precursor to IceCube-Gen2 Radio, it has comparatively smaller effective areas than the latter. This mandates the fiducial choice of $\mathcal{E}_\nu^{\rm UHE,iso} = 2.5\times 10^{52}$ erg, an order of magnitude higher than that chosen for IceCube-Gen2 Radio and extremely optimistic and unlikely. Our results are shown in Fig.~\ref{fig:rnog_tops}. Using RNO-G and CE+ET with the most optimistic $f_{\rm th} = 10^{-2}$, the probability for a coincident UHE neutrino detection is $\sim 10$ years of joint operation, assuming the instantaneous UHE neutrino emission scenario. Constraints at the $2\sigma$ level can be placed in slightly more than $7$ years in the same scenario. If $\delta t = 10^6$ s, 30 years of joint operation time is required to place $2\sigma$ level constraints or to have a $95$\% chance of coincident detection probability. We also show the prospects of RNO-G to constrain the apparent rates assuming optimistic values of energy emitted in UHE neutrinos in Fig.~\ref{fig:rnog_rate_2sig}.
\section{Discussion and outlook}
\label{sec:disc}
In this section we discuss the shortcomings of our current work and the implications. We aim to partially answer the questions we posed in Sec.~\ref{sec:intro}, that is, what could the joint GW and UHE neutrino detectors add to the multimessenger paradigm in the context of searching for UHE neutrinos from BNS mergers.

Let us first discuss the possible improvements and caveats of our current work. Perhaps, the most important shortcoming of this work is the fact that in our analysis, we ignore the trial factors that would be associated with performing a triggered search with an unknown time-window. For each time window we choose to perform our analysis, a weighting needs to be done for the associated trial factor. This is beyond the scope of our current work. The assumption here is that the BNS mergers uniformly emit UHE neutrinos with the same $\delta t$, which is crude. However, this is still useful, since this can enable us to test for and constrain physical models that predict a particular order of magnitude for $\delta t$. Accounting for the trial factors can possibly increase the required timescales for achieving the desired probability of detection or placing constraints in the case of nondetection.

We assume a uniform $E_\nu^{-2}$ UHE neutrino spectrum for this work (see Eq.~\ref{eq:flux}), which serves as a good approximation. Assuming a harder spectra for the neutrinos will lead to a higher fluence and hence improve the detection prospects, and equivalently a softer spectra will make the results less optimistic. Furthermore, $\varepsilon_\nu^{\rm min}$ and $\varepsilon_\nu^{\rm max}$ in Eq.~\eqref{eq:flux} in reality will be determined by the details of the proton or photon spectrum. With enough statistical data, that is, a large sample of UHE neutrinos from BNS mergers as a result of our proposed search technique, one can put constraints on the UHE neutrino fluences and hence in principle can probe the underlying neutrino production processes and the associated spectra. For example, for hadronic production channel of UHE neutrinos in BNS mergers, the constraints obtained can be used to gain insights into key physical quantities like the baryon loading factor, production region, and the details of the proton or photon spectra along with the relevant cut-off energies.

There are other factors that contribute to uncertainties as well. The true fiducial rate has a very large range and can vary between $10\ {\rm Gpc}^{-3}{\rm yr}^{-1}$ to $1700\ {\rm Gpc}^{-3}{\rm yr}^{-1}$. The lower limit would increase the timescale by roughly an order magnitude whereas the upper limits can reduce the same by an order of magnitude. Next, we ignore the duty cycles of the radio neutrino and the GW detectors and assume it to be unity. In reality, the GW detectors have a duty cycle of $\sim 70$\%~\cite{LIGO:2021ppb}. The radio neutrino detectors can generally have a duty cycle of $\sim 100$\%~\cite{GRAND:2018iaj,Gen2_TDR} but the duty-cycle for RNO-G is $\sim 70$\%~\cite{RNO-G:2020rmc}. This can have minor effects in our results. Furthermore, GW events from on-axis jets can have a larger strain and hence a lower associated error region. Since we do not assume this, our results are conservative in this respect. Finally, it is also important to note that we choose $f_{\rm th} = 10^{-3}$ for the combination of CE+ET to make our predictions conservative. A more optimistic prediction can be made by choosing  $f_{\rm th} = 10^{-2}$ for CE+ET. Moreover, the current configurations for both the Einstein Telescope and Cosmic Explorer are uncertain. Depending on how the detectors are oriented their respective sky localization capabilities are subject to change. 

The main theme of this work was to highlight the possibilities that the next-generation GW and the radio neutrino detectors can provide us with, in particular to look for UHE neutrinos from BNS mergers at the radio neutrino detectors. The complementarity of the current and the upcoming radio neutrino detectors is quite powerful in this regard. In Fig.~\ref{fig:fov_nudets} we show the instantaneous field of view (FOV) for GRAND, IceCube-Gen2 Radio, and RNO-G for an arbitrarily chosen date of December 15, 2032 and two different instances during the day corresponding to $12:00:00$ and $23:00:00$. The information on the location and the zenith angle coverage for these detectors is provided in Table~\ref{tab:nu_loc}. In particular, we assume that GRAND will be composed of at least two sub-arrays, for which we show the instantaneous FOV: one in the Northern hemisphere located near the current 300-antenna prototype site, in Xiao Dushan, Gansu, China \cite{2023arXiv230800120G}, and a second in the Southern hemisphere. For the latter, we have chosen as an example, the location of the GRAND@Auger  prototype \cite{2023arXiv230800120G} at the Pierre Auger Observatory, in Argentina, but assuming that the array will be located at an elevation of 1000 m\footnote{The motivation for this site comes in part from the Beamforming Elevated Array for COsmic Neutrinos (BEACON) experiment~\cite{Wissel:2020sec,Southall:2022yil}.}. We see that the instantaneous FOVs help cover large portions in the sky both in the northern ($\delta>0^\circ$) and the southern ($\delta<0^\circ$) hemispheres. This is truly remarkable since the instantaneous FOV is important for short timescale transients in general, and in particular for this work when $\delta t = 100$ s. This can thus ensure that the radio neutrino detectors can collectively have chance to detect UHE neutrinos from mostly all BNS mergers, that is, the mergers are in the combined FOV.

There are a few subtle things associated with the effective areas that are important to note. Firstly, we use the GRAND-200k effective areas. However, if GRAND is built in two different sites, each individual site will have a reduced effective area (by roughly a factor of $1/2$, keeping a constant overall effective area for the full detector) and the FOV of the combined locations would go up by a factor of $2$. Thus one cannot directly map our results assuming GRAND is built in two different sites. 
Using this reasoning, one can make a general remark that for triggered-stacking searches as proposed in this work, the $A_{\rm eff}$ times the FOV is a conserved quantity for various cluster configurations. A narrower FOV may mean that certain events will perhaps be missed. However the deeper effective areas resulting from beamed experiments like GRAND or BEACON, lead to higher rates of neutrino signals. On the other hand, experiments with a broader FOV like IceCube-Gen2 Radio or RNO-G, imply that there is a greater chance that a specific multimessenger event will be seen in the neutrino telescope, although this is not the focus of the current work. Secondly, the apparent difference between the sensitivities of IceCube-Gen2 Radio and GRAND for this analysis is a result of the relevant quantity here being the direction-averaged effective area times the FOV. We stress that for a nearby source this is not true and hence GRAND and IceCube-Gen2 will have different sensitivities and imply different results than what we find from this work. This does not affect our study, since here we focus on far away sources to implement triggered-stacking searches.

Although we only considered joint GW and neutrino observations, EM observations can significantly help improve our results. BNS mergers can be followed by a short GRB or kilonova afterglows. Detection of short GRBs by the satellite Galileo G2~\cite{grb_satellite} or the Space- based multi-band astronomical Variable Objects Monitor (SVOM)~\cite{Wei:2016eox} and x-ray detectors like Swift BAT~\cite{Barthelmy:2005hs}, Einstein Probe~\cite{EinsteinProbeTeam:2015bcj}, and HiZ-GUNDAM~\cite{2020SPIE11444E..2ZY} can help with improving the sky-localization and hence yield higher values of $d_{\rm GW}^{\rm UL}$ which can then improve our results. The kilonova detection horizons for the EM detectors is even more promising. For example, in the optical and infrared bands, the Vera C. Rubin Observatory’s Legacy Survey of Space and Time (LSST)~\cite{Blum:2022dxi,LSST:2008ijt}, Roman~\cite{DES:2017dgt} are $z \sim 0.1$ and $0.2$ respectively and thus can significantly reduce the sky localization areas and lead to more optimistic results than what we present here. In fact we do take this into account in some cases where $d_{\rm GW}^{\rm UL} < 0.5$ Gpc. This is also motivated by the improved second generation of GW detectors like LIGO A+~\cite{KAGRA:2013rdx} and LIGO A\# (A-sharp)~\cite{T2200287}.

The addition of radio neutrino detectors to the existing GW, neutrino, and EM detectors will bolster the search strategies for UHE neutrino events from BNS mergers. The implications of a coincident UHE neutrino event from a BNS merger or constraining the parameter space at $2\sigma$ or $3\sigma$ C.L.s in the case of nondetections, in reasonably short timescales of joint operation are major. Our current work presents an effective way and elaborates on a search technique that is timely and useful for the upcoming era of multimessenger astrophysics, particularly in the context of BNS mergers and stacking searches from transient sources.
\acknowledgements
We thank Rafael Alves Batista, Mauricio Bustamante, Cosmin Deaconu, Christian Glaser, Kaeli Hughes, Marco Stein Muzio, and Foteini Oikonomou for comments and useful discussions. We are grateful to Christian Glaser for providing us with the effective volumes for IceCube-Gen2 Radio, Foteini Oikonomou and the GRAND Collaboration for providing the effective areas for GRAND. We also thank Pablo Correa for carefully reading our manuscript on behalf of the GRAND collaboration.
M.\,M. and K.\,M. are supported by NSF Grant No. AST-2108466. M.\,M. also acknowledges support from the 
Institute for Gravitation and the Cosmos (IGC) Postdoctoral Fellowship. 
K.\,K. acknowledges support from the Fulbright-France program, the CNRS Programme Blanc MITI ("GRAND" 2023.1 268448; France), and the CNRS Programme AMORCE ("GRAND" 258540; France). 
S.W. acknowledges support from NSF Grant Nos.~2033500, 2111232, and 2310122.
The work of K.M. is supported by the NSF Grant Nos.~AST-2108467, and AST-2308021, and KAKENHI Nos.~20H01901 and 20H05852. S.S.K. acknowledges the support by KAKENHI No. 22K14028, No. 21H04487, No. 23H04899, and the Tohoku Initiative for Fostering Global Researchers for Interdisciplinary Sciences (TI-FRIS) of MEXT’s Strategic Professional Development Program for Young Researchers.
\appendix
\section{Effective volumes and effective areas}
\label{appsec:effarea}
Unlike optical neutrino telescopes like IceCube or KM3NeT, radio neutrino detectors are sensitive to cascades $\gtrsim 10^{16}$ eV. The prospects for detection of a cascade thus depends on the orientation of the Cherenkov cone. Besides, the sparse configuration of the detectors necessitates the use of effective volumes instead effective areas. To achieve this the effective volumes $\mathcal{V}_{\rm eff}(E_\nu, \theta_z)$ are computed by simulating neutrino interactions for a given neutrino energy $E_\nu$ and zenith angle $\theta_z$ in a volume $V$. Assuming a thin-target this effective volume is converted to effective area $\mathcal{A}_{\rm eff} (E_\nu, \theta_z)$
\be
\mathcal{A}_{\rm eff} (E_\nu, \theta_z) = \frac{\mathcal{V}_{\rm eff}(E_\nu, \theta_z)}{\mathcal{L}_{\rm int}(E_\nu)}\,,
\ee
where the interaction length of a neutrino in the Earth is defined as $\mathcal{L}_{\rm int}(E_\nu)$, which depends on cross-sections which in turn depends on the neutrino energies.

There are different ways to estimate an effective area for a given neutrino detector
\begin{itemize}
\item \emph{instantaneous effective area} - which is relevant for very short periods of time $\sim 1 - 1000$ s,
\item \emph{day-averaged effective area} - which is relevant searching for neutrinos from long transients $\sim 10^5 - 10^7$ s,
\item \emph{direction-dependent effective area} - which can depend on the right ascension (RA) and/or the declination (dec),
\item  \emph{direction-averaged effective area} - which is computed by integrating over the complete solid angle.
\end{itemize}
IceCube-Gen2 Radio's location at the south pole makes it such that the instantaneous and the day-averaged effective area is the same. However, the same is not true for GRAND and RNO-G. In this work, we use the zenith (or declination)-dependent effective area for each of the neutrino detectors.

Assume a radio neutrino detector has an effective area which is sensitive between zenith angles $\theta_z = [\theta_z^{\rm min},\theta_z^{\rm max}]$ at any instant. The direction-averaged effective area can be defined as,
\be
\label{appeq:aeff}
\mathcal{A}_{\rm eff}^{\rm avg.} = \frac{1}{N_{\rm norm}} \int d\Omega \mathcal{A}_{\rm eff}(\phi,\theta_z,E_\nu)\,,
\ee
where $N_{\rm norm}$ is the normalization, $\mathcal{A}_{\rm eff}(\phi,\theta_z,E_\nu)$ is the instantaneous azimuth, zenith\footnote{The zenith angle $\theta_z$ can be converted to altitude angle $\alpha$ by $\alpha = 90^\circ - \theta_z$. However conversion to declination angle $\delta$ is non-trivial and depends on the location (latitude and longitude) of the detector and the time of the day.}, and neutrino energy dependent effective area, $\phi$ is the azimuthal angle and $\phi = [0,2 \pi]$ and $d\Omega$ is the integration over solid angle given as $d\Omega = d\phi\ d({\rm cos}\theta_z)$. Since GRAND, IceCube-Gen2 Radio, and RNO-G have sensitivities to the whole azimuth at any instant and $\mathcal{A}(\phi,\theta_z,E_\nu) = \mathcal{A}(\theta_z,E_\nu)$. Thus the integration over $d\phi$ is trivial and gives a factor of $2\pi$. The choice of $N_{\rm norm}$ will depend on whether we want to average over all sky, that is, $4\pi$ or over the field of view of the detector, in which case, $N_{\rm norm} = 2 \pi \left| \big( {\rm cos}\ \theta_z^{\rm min} - {\rm cos}\ \theta_z^{\rm min} \big) \right|$. Note that when the full zenith is covered, as is the case for IceCube-Gen2 $N_{\rm norm} = 4\pi$, which gives the all-sky average. In other words, Eq.~\eqref{appeq:aeff} provides us with the direction-averaged neutrino energy dependent effective area in a given zenith band.

For the purposes of triggered searches from distances $\gtrsim 500$ Mpc, the number of neutrinos in the detector $N_\nu << 1$ (see Eq.~\ref{eq:nnu}). In this scenario Eq.~\eqref{eq:totprob} can be re-written as
\be
\label{appeq:prob}
P_{n\geq 1} (d_L) = \frac{1}{N_{\rm norm}} \int d\Omega\ N_\nu(\theta_z,d_L)\,.
\ee
The field of view (FOV) for the detector can be defined as the fraction of total sky area covered by the detector and can be given as
\be
{\rm FOV} =  \frac{2 \pi \left| \big( {\rm cos}\ \theta_z^{\rm min} - {\rm cos}\ \theta_z^{\rm min} \big) \right|}{4 \pi}\,.
\ee
The radio neutrino detector will only detect a neutrino if it is in its FOV. Thus the number of sources in the FOV is given by 
\be
N_{\rm src} = {\rm FOV} \int_0^{d^\prime_{\rm com}} d(d_{\rm com}) \big( 4\pi d_{\rm com}^2 R_{\rm app}(z) \big)\,.
\ee
Using Eqns.~\eqref{appeq:aeff} and~\eqref{appeq:prob}, Eq.~\eqref{eq:qtop} can be re-written as
\begin{widetext}
\centering
\begin{align}
\label{appeq:qtop}
q\big( d^{\rm UL}_{\rm GW}, T_{\rm op} \big) &= 1 - {\rm exp}\bigg( -T_{\rm op} I \big( d_{\rm GW}^{\rm UL} \big) \bigg) \nonumber\,,\\
I \big( d_{\rm GW}^{\rm UL} \big) = 4\pi \int_{0}^{d_{\rm GW}^{\rm UL}} d (&d_{\rm com}) \frac{T_{\rm op}}{\big( 1+z \big)} \left[ \frac{2 \pi \big( {\rm cos}\ \theta_z^{\rm min} - {\rm cos}\ \theta_z^{\rm min} \big)}{4 \pi} \right] R_{\rm app}\big( z \big) d_{\rm com}^2 \nonumber\\ 
&\left[ \frac{2\pi \int_{E_{\nu}^{\rm LL}}^{E_{\nu}^{\rm UL}} dE_{\nu} \phi_\nu \big(\mathcal{E}_\nu^{\rm UHE,iso}, E_\nu ,d_L \big) \int d({\rm cos}\theta_z) \mathcal{A}_{\rm eff}(\theta_z,E_\nu)}{2 \pi \big( {\rm cos}\ \theta_z^{\rm min} - {\rm cos}\ \theta_z^{\rm min} \big)}  \right]\,,\\
I \big( d_{\rm GW}^{\rm UL} \big) = 4\pi \int_{0}^{d_{\rm GW}^{\rm UL}} d (&d_{\rm com}) \frac{T_{\rm op}}{\big( 1+z \big)} R_{\rm app}\big( z \big) d_{\rm com}^2
\left[ \int_{E_{\nu}^{\rm LL}}^{E_{\nu}^{\rm UL}} dE_{\nu} \phi_\nu \big(\mathcal{E}_\nu^{\rm UHE,iso}, E_\nu ,d_L \big)  \frac{1}{2} \int d({\rm cos}\theta_z) \mathcal{A}_{\rm eff}(\theta_z,E_\nu)  \right] \nonumber\,.
\end{align}
\end{widetext}
The above equation is universally true for triggered-searches when the sources are very far away, that is, $N_\nu << 1$, which is what we consider in this paper. Recall that $\mathcal{A}_{\rm eff} (\theta_z, E_\nu)$ is the instantaneous zenith-dependent effective area. As evident from above Eq.~\eqref{appeq:qtop}, the results depend \emph{only} on the \emph{instantaneous zenith-dependent effective area} $\mathcal{A}_{\rm eff} (\theta_z, E_\nu)$, no matter the size of the triggered-search time window ($\delta t$). This is because in the above equations the product of $\mathcal{A}_{\rm eff} (\theta_z, E_\nu) \times {\rm FOV}$ is a constant and is equal to $(1/2) \int d({\rm cos}\theta_z) \mathcal{A}_{\rm eff}(\theta_z,E_\nu)$. This highlights the fact that larger instantaneous effective areas will lead to better results irrespective of the FOV for this analysis and hence explains why GRAND yields better results than IceCube-Gen2 Radio.
\bibstyle{aps}
\bibliography{refs}
\end{document}